\documentclass[
pre,notitlepage
]{revtex4-1}

\usepackage{graphicx}
\usepackage{dcolumn}
\usepackage{bm}
\usepackage{indentfirst}

\usepackage{amsmath, amssymb, amsthm}
\usepackage{epsfig, color}

\begin{document}

\title{Augmented Variational Superposed Gaussian Approximation for Langevin Equations with Rational Polynomial Functions}

\author{Xingyan Chu}
\email{chu@biom.t.u-tokyo.ac.jp}
\affiliation{
	Department of Information and Communication Engineering, Graduate School of Information Science and Technology,\\
	The University of Tokyo, Tokyo 113-8656, Japan
}

\author{Yoshihiko Hasegawa}
\email{hasegawa@biom.t.u-tokyo.ac.jp}
\affiliation{
	Department of Information and Communication Engineering, Graduate School of Information Science and Technology,\\
	The University of Tokyo, Tokyo 113-8656, Japan
}

\date{\today}

\begin{abstract}
Reliable methods for obtaining time-dependent solutions of Langevin equations are in high demand in the field of non-equilibrium theory.
In this paper, we present a new method based on variational superposed Gaussian approximation (VSGA) and Pad\'e approximant.
The VSGA obtains time-dependent probability density functions as a superposition of multiple
Gaussian distributions.
However, a limitation of the VSGA is that the expectation of the drift term
with respect to the Gaussian distribution should be calculated analytically,
which is typically satisfied when the drift term is a polynomial function. 
When this condition is not met, the VSGA must rely on the numerical integration of the expectation
at each step, resulting in huge computational cost.
We propose an augmented VSGA (A-VSGA) method that effectively overcomes the limitation of the VSGA by approximating non-linear functions with the Pad\'e approximant.
We apply the A-VSGA to two systems driven by chaotic input signals, a stochastic genetic regulatory system and a soft bistable system, whose drift terms are a rational polynomial function and a hyperbolic tangent function, respectively. The numerical calculations show that the proposed method can provide accurate results with less temporal cost than that required for Monte Carlo simulation.

\end{abstract}
\maketitle

\section{Introduction\label{sec:introduction}}
In recent years, non-equilibrium systems have attracted much interest by virtue of advancements in non-equilibrium thermodynamics \cite{ritort2008review, seifert2012stochastic}. To investigate the dynamics of non-equilibrium systems, studies have focused on the properties and analytical solutions of Langevin equations \cite{hida1980brownian,van1992stochastic,coffey2004langevin,fa2011solution}.
For stationary systems, the solutions of Langevin equations do not depend on time and can be calculated analytically for one-dimensional systems.
However, systems are driven by external forces in dynamic cases.
Except for some particular cases, time-dependent solutions cannot be obtained analytically for such driven systems, not even for one-dimensional cases \cite{risken1996fokker}.
Therefore, the time-dependent solutions of Langevin equations with external forces or input signals remain a difficult problem.

Our goal is to propose a reliable numerical method that can be used to solve time-dependent solutions of Langevin equations with driving forces to meet the needs of non-equilibrium theory \cite{ritort2008review, seifert2012stochastic,weeks2007advances}.
The moment method is a widely applied approach which computes the time evolution of the mean and the covariance of Langevin equations \cite{rodriguez1996spikeNeurons, tuckwell2009HHeq}.
Although it is simple and computationally efficient, 
its accuracy is not satisfactory even for simple bistable models.
Direct numerical approaches such as a finite-element
method \cite{harrison1988fem, kumar2006fem} and a finite-difference method \cite{whitney1970fdm} were proposed
to obtain time-dependent probability density functions (PDFs), but the computational cost associated with the methods was very high.
Considering the success of the multiple Gaussian method with the variational principle for time-dependent wave functions in quantum mechanics \cite{heller1976time}, several studies have tried to approximate solutions of Langevin equations by a superposition of Gaussian distributions \cite{er1998multi,pradlwarter2001non,terejanu2008uncertainty,hasegawa2015variational}.
Er~\cite{er1998multi} obtained PDFs as a
superposition of multiple Gaussian distributions with the weighted
residual method, but the approach is limited to stationary cases. 
Pradlwarter~\cite{pradlwarter2001non} represented 
time-dependent solutions by a sum of small
elements of Gaussian distributions. A disadvantage of this method is that it must handle the variance and the number
of Gaussian distributions. 
Terejanu \emph{et al.} \cite {terejanu2008uncertainty} proposed a superposition of Gaussian distributions
as time-dependent PDFs. 
Their approach first assumed that the weights for each of the Gaussian distributions are fixed. 
After calculating the mean and variance of the Gaussian distributions, the weights are optimized by minimizing the squared error. The variational superposed Gaussian approximation (VSGA) is shown to be effective and accurate in the calculation of time-dependent solutions in strongly non-linear systems \cite{hasegawa2015variational}.
When compared with \cite{er1998multi,terejanu2008uncertainty,pradlwarter2001non}, the VSGA directly obtains the time evolution of the mean, variance, and weight and does not require extra steps.
Still, a drawback of the VSGA is that the expectation of the drift term with respect to the Gaussian distribution should be obtained analytically (e.g., when the drift term is a polynomial function). Otherwise, the expectation would be calculated by time-consuming numerical integration.
This requirement greatly restricts the application of the VSGA.

In this paper, we propose an augmented VSGA (A-VSGA) method that can handle systems with rational polynomial functions.
Rational functions can be used to represent many dynamical systems such as biochemical reactions, gene regulation, and dose-response.
Furthermore, through Pad\'e approximant, non-linear systems with drift terms (irrational functions) can be approximated by rational functions. The use of Pad\'e approximant enables applications of the A-VSGA to such non-linear systems.
In many cases, the Pad\'e approximant provides closer approximation than the Taylor series expansion, and it works even when the Taylor expansion diverges at poles.
To test the accuracy and effectiveness of the A-VSGA, we first apply the A-VSGA to a stochastic gene regulatory system where the drift term is represented by a Hill equation.
The Hill equation is a fundamental equation to describing the cooperative effects of multimers. It has been widely used in systems biology \cite{weiss1997hill,tian2010stochastic,smolen2000modeling,liu2004fluctuations}.
Next, to investigate its reliability, we apply the A-VSGA with the Pad\'e approximant to a soft bistable system subject to white Gaussian noise.
The soft bistable model reflects the dynamics of the membrane potential of a neuron cell and is represented by a hyperbolic tangent function \cite{longtin1994bistability}.
As noted above, its drift term can be approximated by the rational polynomial functions via the Pad\'e approximant.
Bistability describes the switching dynamics of systems and often occurs in non-linear systems far from equilibrium \cite{wilhelm2009smallest}. Due to non-linearity, many researches have studied bistable systems by simulation only. We demonstrate our method as an effective and accurate approach for solving time-dependent solutions of Langevin equations with a non-polynomial expression of the drift terms.

The paper is organized as follows. 
In Sec.~\ref{sec:method}, we briefly introduce the VSGA and its shortcomings, and propose an augmented method, the A-VSGA.
Then, we investigate the accuracy and effectiveness of the A-VSGA in Sec.~\ref{sec:results} by applying it to a gene regulatory system and a soft bistable system driven by time-dependent input signals. 
Finally, we discuss the problems in the experiments and summarize our findings in Sec.~\ref{sec:conclusion}.

\section{Methods\label{sec:method}}

An $N$-dimensional Langevin equation, 
\begin{equation}
\frac{dx_i}{dt}=f_i(\bm{x},t)+\sum\limits_{j=1}^{N_g}g_{ij}(\bm{x},t)\xi_j(t),\quad(i=1,2,...,N),
\label{eq:Langevin_def} 
\end{equation}
describes the temporal evolution of $N$ state variables $\bm{x}=(x_1,x_2,...,x_N)^\top$. 
In Eq.~\eqref{eq:Langevin_def}, $f_i(\bm{x},t)$ and $g_{ij}(\bm{x},t)$ denote drift and multiplicative terms, respectively. $\xi_i(t)$ is white Gaussian noise with zero mean $\left\langle\xi_i(t)\right\rangle=0$, and the correlation is $\left\langle\xi_i(t)\xi_j(t')\right\rangle=2\delta_{ij}\delta(t-t')$, where $\delta$ is the Dirac delta function.
We interpret Eq.~\eqref{eq:Langevin_def} in the Stratonovich sense. At a given time $t$, the state of system is not deterministic due to the noise, $\bm{\xi}(t)$. 
By repeatedly solving Eq.~\eqref{eq:Langevin_def} for all $N$ states of $\bm{x}$, the probability density functions (PDFs), $P(\bm{x};t)$, can be obtained numerically.
Since the focus of this research is the PDFs of Langevin equations with external forces, the Fokker-Planck equation (FPE), which gives the time evolution of the PDF, is used and is given by Eq.~\eqref{eq:FPE_def} \cite{risken1996fokker},
\begin{equation}
    \frac{\partial}{\partial t}P(\bm{x};t)=\widehat{L}(\bm{x};t)P(\bm{x};t), \label{eq:FPE_def}
\end{equation}
where $P(\bm{x},t)$ is the PDF of $\bm{x}$ at time $t$, and $\widehat{L}(\bm{x};t)$ is an FPE operator written as
\begin{equation}
    \widehat{L}(\bm{x};t)=-\sum\limits _{i}\frac{\partial}{\partial x_{i}}\left[f_{j}(\bm{x})+\sum\limits _{k,j}g_{kj}(\bm{x};t)\frac{\partial}{\partial x_{k}}g_{ij}(\bm{x};t)\right]+\sum\limits _{i,j}\frac{\partial^{2}}{\partial x_{i}\partial x_{j}}\sum\limits _{k}g_{ik}(\bm{x};t)g_{jk}(\bm{x};t).
    \label{eq:FPE_op_def}
\end{equation}
The time-dependent PDF can be obtained by solving Eq.~\eqref{eq:FPE_def}.

\subsection{Variational Superposed Gaussian Approximation (VSGA)\label{sec:introduction to Variational superposed Gaussian approximation}}

The VSGA is an adaptation of the multiple Gaussian wave packets approximation in quantum mechanics to Langevin equations \cite{skodje1984multiackets, worth2004multigwp, zoppe2005gpackets}.
It uses superposed multiple Gaussian distributions to approximate PDFs and obtain time-evolution equations for parameters of each Gaussian distribution with the variational principle \cite{hasegawa2015variational}.

We first review procedures of the VSGA (for details of the method, please see \cite{hasegawa2015variational}).
Starting from the FPE, Eq.~\eqref{eq:FPE_def}, let $\Theta(\bm{x};t)$ be the time derivative of $P(\bm{x};t)$.
Since $\Theta(\bm{x};t) dt + P(\bm{x};t) = P(\bm{x};t+dt)$, 
given $P(\bm{x};t)$ at time $t$, $P(\bm{x};t+dt)$ is specified when $\Theta(\bm{x};t)$ is determined. 
Thus we seek to obtain the optimal $\Theta(\bm{x};t)$ such that $P(\bm{x};t+dt)$ 
optimally satisfies the FPE (Eq.~\eqref{eq:FPE_def}) with given $P(\bm{x};t)$.
The optimal $\Theta(\bm{x};t)$ should minimize
\begin{equation}
    R[\Theta]=\int_{-\infty}^{\infty}\left[\widehat{L}(\bm{x};t)P(\bm{x};t)-\Theta(\bm{x};t)\right]^2 d\bm{x}.
    \label{eq:R_Theta}
\end{equation}
When $P(\bm{x};t)$ satisfies Eq.~\eqref{eq:FPE_def}, $R[\Theta]$ always vanishes. Ensuring the normalization condition ($\int P(\bm{x};t)d\bm{x} = 1$) of the PDF yields a constraint condition on $\Theta(\bm{x};t)$ as $\int_{-\infty}^{\infty}\Theta(\bm{x};t)d\bm{x}=0$. 
To minimize Eq.~\eqref{eq:R_Theta} with the normalization condition equation, we introduce the Lagrange multiplier $\lambda(t)$ as follows:
\begin{equation}
    \widetilde R[\Theta]=\int_{-\infty}^{\infty}{\left[\widehat{L}(\bm{x};t)P(\bm{x};t)-\Theta(\bm{x};t)\right]}^2 d\bm{x} +\lambda(t)\int_{-\infty}^{\infty}\Theta(\bm{x};t)d\bm{x}.\label{eq:R_tilde}
\end{equation}
With the variation of $\delta\Theta(\bm{x};t)$ in Eq.~\eqref{eq:R_tilde}, Eq.~\eqref{eq:R_tilde_vanish} should hold for optimal $\Theta(\bm{x};t)$,
\begin{equation}
    \int_{-\infty}^{\infty}\delta\Theta\left[\widehat{L}(\bm{x};t)P(\bm{x};t)-\Theta(\bm{x};t)+\lambda(t)\right]d\bm{x}=0.
    \label{eq:R_tilde_vanish}
\end{equation}
The approximate $P(\bm{x};t)$ obtained with the superposition of multiple Gaussian distributions is given as follows:
\begin{equation}
    P(\bm{x};t)=P(\bm{x};\bm{\theta}(t))=\sum_{m=1}^{N_B} r_m(t) \exp\left[-\bm{x}^\top\mathbf{A}_m(t)\bm{x}+\bm{b}^\top_m(t)\bm{x}\right],\label{eq:superposed_Gaussian}
\end{equation}
where $\mathbf{A}_m$ is a positive definite $N\times N$ symmetric matrix, $\bm{b}_m$ is an $N$-dimensional column vector, and $r_m$ is the weight of the $m$th basis function. $N_B$ is the number of basis functions.
In Eq.~\eqref{eq:superposed_Gaussian}, a parameter set $\bm{\theta}(t)=(\theta_1(t),\theta_2(t),...,\theta_K(t))$ is composed of $\mathbf{A}_m$, $\bm{b}_m$, and $\bm{r}_m$.  
For an $N$-dimensional system and $N_B$ basis functions, the total number of parameters is 
\begin{equation}
    K=\frac{N_B(N+1)(N+2)}{2}.\label{eq:K_def}
\end{equation}
Thus, Eq.~\eqref{eq:R_tilde_vanish} can be written as follows:
\begin{equation}
    \int_{-\infty}^{\infty}\frac{\partial P(\bm{x};\bm{\theta})}{\partial \theta_l}\left[\widehat{L}(\bm{x};t)P(\bm{x};\bm{\theta})-\Theta(\bm{x};\bm{\theta},\dot{\bm{\theta}})+\lambda(t)\right]d\bm{x}=0,\quad(l=1,2,...,K).\label{eq:DAE_solution}
\end{equation}
We can uniquely obtain parameters of multiple Gaussian distributions by solving the $K$ constraints (Eq.~\eqref{eq:DAE_solution}) and ensuring a normalization condition.

\subsection{Augmented Variational Superposed Gaussian Approximation (A-VSGA)}\label{sec:our method}

As reported in \cite{hasegawa2015variational}, the VSGA provides very accurate approximations for one- and two-dimensional driven systems. 
Still, this approach is limited in some respects; the integral in Eq.~\eqref{eq:DAE_solution} can be computed analytically if 
the expectation of the drift term allowing the Gaussian integration, which enables an efficient calculation of time-dependent solutions.
Typically, this condition is met when the drift term is given by polynomial functions because the expectation with respect to Gaussian distribution can be calculated analytically for arbitrary degrees of polynomials. 
However, the potentials of most non-linear models are not polynomial \cite{de2013non}, and this makes the VSGA computationally expensive. Consequently, in the A-VSGA, we mitigate that restriction so that the approximation can handle rational polynomial functions.

We explain our method by reviewing a one-dimensional case with an additive noise term for simplicity.
Essentially, the same calculation can be conducted for multivariate cases and for systems with multiplicative noise terms given by rational functions. We show that Eq.~\eqref{eq:DAE_solution} can be solved when the drift term is a rational polynomial function. Suppose that the drift function is given by the rational function
\begin{equation}
    f(x)=\frac{g(x)}{h(x)},\label{eq:poly_rat_def}
\end{equation}
where $g(x)$ and $h(x)$ are polynomials.
Substituting Eq.~\eqref{eq:poly_rat_def} into Eq.~\eqref{eq:FPE_def}, we obtain
\begin{equation}
    \frac{\partial}{\partial t}P(x;t)=\Theta(x;t)=-\frac{\partial}{\partial x}\frac{g(x)}{h(x)}P(x;t)+D\frac{\partial^2}{\partial x^2}P(x;t).\label{eq:FPE_ratpoly}
\end{equation}
Multiplying Eq.~\eqref{eq:FPE_ratpoly} by $h^2(x)$, we find
\begin{equation}
    0=-h^2(x)\Theta(x;t) -\left[\frac{\partial g(x)P(x;t)}{\partial x}h(x)-\frac{\partial h(x)}{\partial x}g(x)P(x;t)\right]+Dh^2(x)\frac{\partial^2}{\partial x^2}P(x;t).\label{eq:FPE_ratpoly2}
\end{equation}
When the time evolution of $P(x;t)$ satisfies Eq.~\eqref{eq:FPE_def} with the drift term (Eq.~\eqref{eq:poly_rat_def}), the right-hand side of Eq.~\eqref{eq:FPE_ratpoly2} should vanish.
Therefore, the optimal $\Theta(x;t)$ should minimize the following performance index:
\begin{equation}
    R[\Theta]=\int_{-\infty}^{\infty}\left[-\frac{\partial g(x)P(x;t)}{\partial x}h(x)+\frac{\partial h(x)}{\partial x}g(x)P(x;t)+Dh^{2}(x)\frac{\partial^{2}}{\partial x^{2}}P(x;t)-h^{2}(x)\Theta(x;t)\right]^{2}dx.\label{R_ratpoly_def}
\end{equation}
In a similar procedure to that of the VSGA, the Lagrange multiplier $\lambda(t)$ for the normalization constraint is introduced, and the following $3N_B$ equations are obtained (as Eq.~\eqref{eq:K_def} implies, the total number of parameters is $K=3N_B$ for one-dimensional systems):
\begin{equation}
    \int_{-\infty}^{\infty}\frac{\partial P(x;\theta)}{\partial\theta_{l}}[\widehat{L}(x;t)P(x;\theta)-h^{4}(x)\Theta(x;\theta,\dot{\theta})+\lambda(t)]dx=0,\hspace*{1em}(l=1,2,\dots,3N_{B}),\label{eq:ratpoly_coupled_equations}
\end{equation}
where
 $$\widehat{L}(x;t)P(x;\theta)=\frac{\partial g(x)P(x;\theta)}{\partial x}h^3(x)-\frac{\partial h(x)}{\partial x}g(x)P(x;\theta)h^2(x)+D h^4(x) \frac{\partial^2}{\partial x^2}P(x,\theta)).$$

We show that the A-VSGA can handle rational polynomial systems. Furthermore, it can be used to approximate non-linear systems with irrational potential functions. Our idea is that the drift terms can be approximated by a rational function via methods such as Taylor series expansion and the Pad\'e approximant. Although Taylor expansion is an expression of functions as an infinite sum of monomials, it may not converge over the entire domain. 
Even in such a situation, the Pad\'e approximant still works and often yields better approximation than truncating the Taylor series \cite{stewart1960generalized,fair1964pade,lorentz1996constructive,cohen1991pade}.
A given function $V(x)$, can be approximated by the Pad\'e approximant of order $[m/n]$ as follows:
\begin{equation}
    V(x)\approx\frac{a_0+a_1x+a_2x^2+...+a_mx^m}{1+b_1x+b_2x^2+..+b_nx^n}=\frac{g(x)}{h(x)},\quad(m\ge0, n\le1, m,n\ \text{are integers}).\label{eq:Pade_explanation}
\end{equation}
For systems with irrational potential, we first convert the drift terms to rational functions by using Pad\'e approximants of a suitable order. From Eq.~\eqref{eq:ratpoly_coupled_equations} and the normalization condition, we have $3N_B+1$ implicit differential equations. By numerically solving these differential equations (\emph{Mathematica 10} is used for the implementation), we can obtain accurate time-dependent solutions.

\section{Results\label{sec:results}}
We apply the A-VSGA to a stochastic biochemical reaction and a soft bistable model. 
We also conduct Monte Carlo (MC) simulations to confirm the reliability of the A-VSGA.
For MC simulations, we employ the Euler method with time resolution $\Delta t = 0.001$, and the PDFs were constructed by repeating the stochastic simulations $100 000$ times.

\subsection{Gene Regulatory System}

\begin{figure}[t]
\includegraphics[width=6cm]{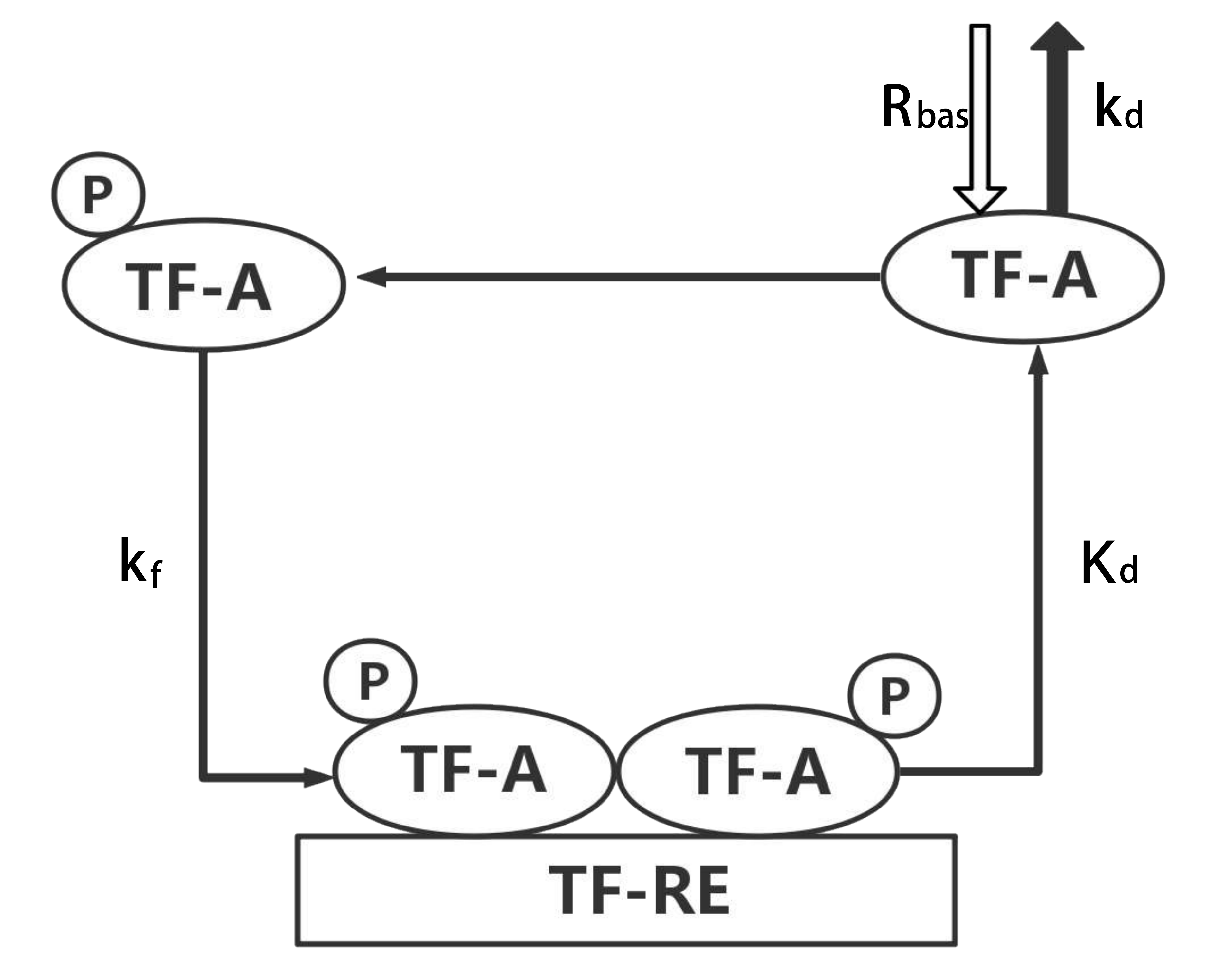}
\caption{Model of genetic regulation with a positive autoregulatory feedback loop. When phosphorylated (circled ``P'' denotes the phosphorylation) TF-A homodimers bind to TF-RE, TF-A transcription is increased. $k_f$ is the maximal rate of the TF-A that activates transcription. TF-A is degraded with rate $k_d$ and synthesized with rate $R_{\mathrm{bas}}$. \label{fig:TF}}
\end{figure}

We first apply the A-VSGA to a model of genetic regulation with a positive auto-regulatory feedback loop. A flow diagram of the model is found in FIG.~\ref{fig:TF}. The transcription factor activator (TF-A) gene incorporates a transcription factor responsive element (TF-RE). When phosphorylated TF-A homodimers bind to this element, the transcription is increased. The fraction of dimers phosphorylated is dependent on the activity of kinases whose activity can be regulated by external signals \cite{liu2004fluctuations} (for more details of the model, please see \cite{liu2004fluctuations}). The change rate of concentration of the TF-A dimer is described by the following Hill equation:
\begin{equation}
    \frac{dx}{dt}=\frac{k_fx^2}{x^2+K_d}-k_dx+R_\mathrm{bas}, \label{eq:hill equation}
\end{equation}
where $k_f$ is the maximal rate of TF-A that activates transcription. TF-A is degraded at a rate of $k_d$ and synthesized at a rate of $R_\mathrm{bas}$. $K_d$ is the dissociation concentration of the TF-A dimer from TF-REs.

To simulate the stochastic effects of the synthesis process, we assume that white Gaussian noise, $\xi(t)$, is added to the reaction rate as $R_\mathrm{bas}\rightarrow R_\mathrm{bas}+\xi(t)$. Thus, Eq.~\eqref{eq:hill equation} becomes the following Langevin equation:
\begin{equation}
    \frac{dx}{dt}=\frac{k_fx^2}{x^2+K_d}-k_dx+I(t)+R_\mathrm{bas}+\sqrt{D}\xi(t), \label{eq:Langevin_Hill equation}
\end{equation}
where $D$ is the noise intensity, $I(t)$ is an external driving signal, and $\xi(t)$ is white Gaussian noise. We set the parameters to $k_f=6$, $k_d=1$, $K_d=10$, and $R_\mathrm{bas}=0.4$, as is done in \cite{liu2004fluctuations}, to make the system bistable.
The corresponding FPE operator $\widehat{L}(x;t)$ is given by Eq.~\eqref{eq:FPE_op_Hill},
\begin{equation}
    \widehat{L}(x;t)=-\frac{\partial}{\partial x}\left[\frac{k_fx^2}{x^2+K_d}-k_dx+I(t)+R_\mathrm{bas}\right]+D\frac{\partial^{2}}{\partial x^{2}}.\label{eq:FPE_op_Hill}
\end{equation}
Substituting Eq.~\eqref{eq:FPE_op_Hill} into Eq.~\eqref{eq:ratpoly_coupled_equations}, we have $3N_B+1$ constraints including a normalization condition.

We first consider the stationary case, i.e., $I(t)$ is constant, since stationary PDFs can be calculated analytically for it. The analytic solution is given by Eq.~\eqref{eq:PDF_analytic} \cite{er1998multi},
\begin{equation}
    P_{st}(x)=\frac{1}{Z(D)}\exp{\left[-\frac{U(x)}{D}\right]},
\label{eq:PDF_analytic}
\end{equation}
where $Z(D)=\int_{-\infty}^{\infty}\exp{\left[-U(x)/D\right]} dx$ and $U(x)=-\int\left[k_fx^2/\left(x^2+K_d\right)-k_dx+R_\mathrm{bas}\right]dx=-k_fx-R_\mathrm{bas}x+k_dx^2/2+k_f\sqrt{K_d}\arctan{\left(x/\sqrt{K_d}\right)}$.

\begin{figure}[t]
\centering
\includegraphics[width=0.325\linewidth]{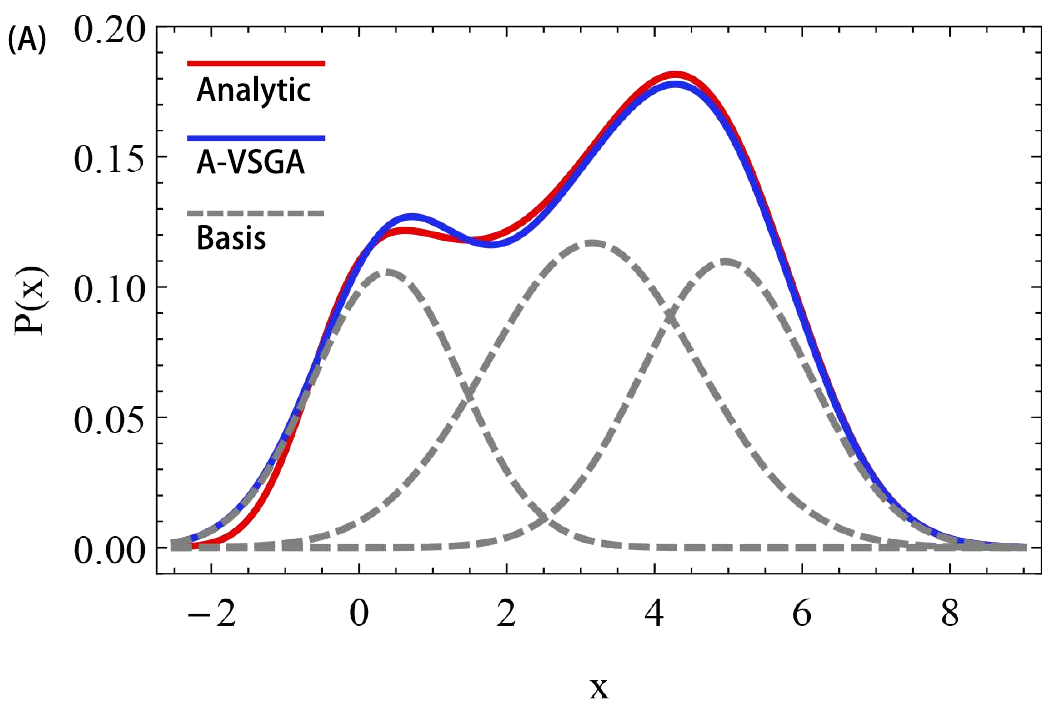}\label{main:hill_a}
\includegraphics[width=0.325\linewidth]{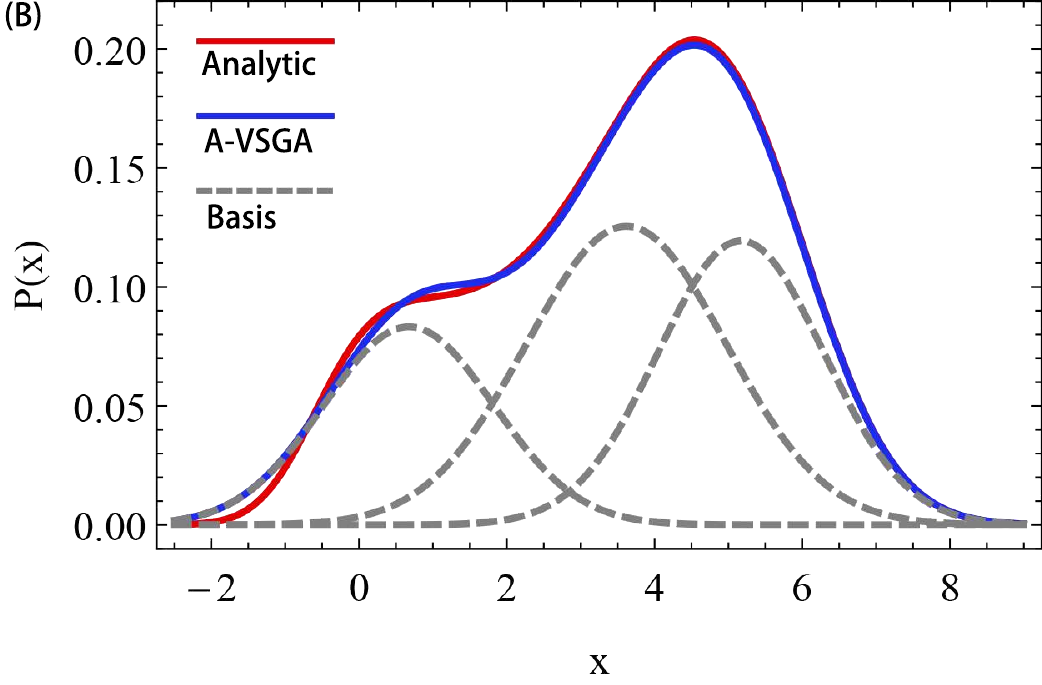}\label{main:hill_b}
\includegraphics[width=0.325\linewidth]{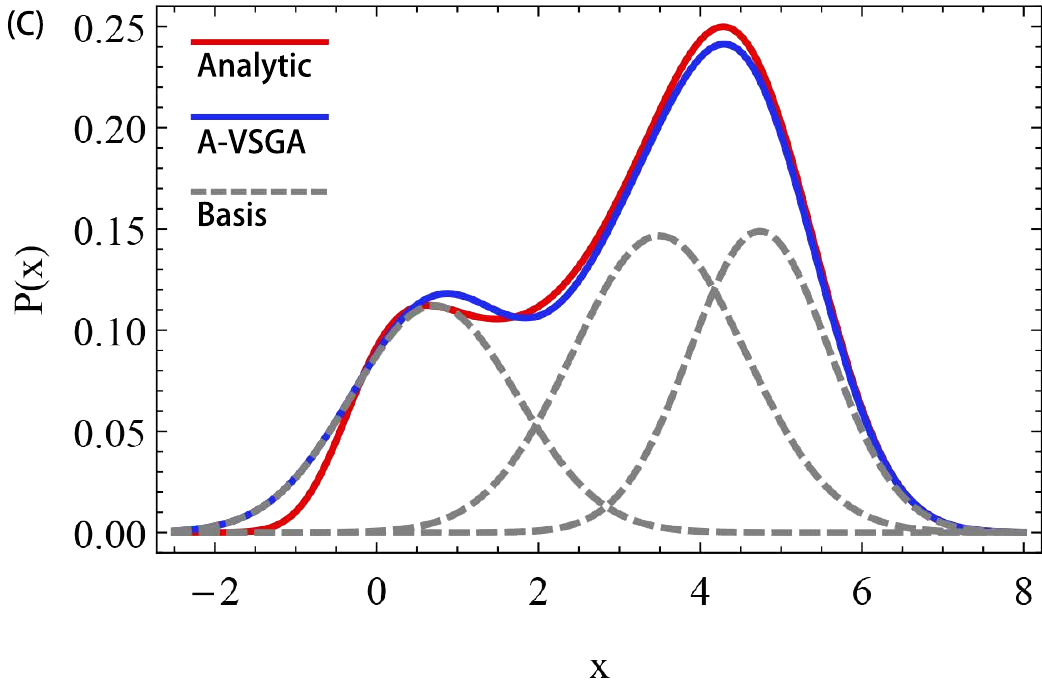}\label{main:hill_c}
\caption{(Color online) Stationary PDFs of a gene regulatory system obtained  by analytic calculation of Eq.~\eqref{eq:PDF_analytic} (red solid line) and the A-VSGA (blue solid line) for (A) $D=1.0$, $I(t)=0$, and $N_B=3$,  (B) $D=1.0$, $I(t)=0.1$, and $N_B=3$, and (C) $D=0.5$, $I(t)=0$, and $N_B=3$. The grey dashed lines represent the Gaussian basis functions. }
\label{fig:station_Hill}
\end{figure}

FIG.~\ref{fig:station_Hill} shows the stationary solutions obtained by the A-VSGA (blue line) and the analytic solutions (red line) obtained by Eq.~\eqref{eq:PDF_analytic} for (A) $D=1.0, I(t)=0, N_B=3$; (B) $D=1.0,I(t)=0.1,N_B=3$; and (C) $D=0.5, I(t)=0, N_B=3$. 
Note that there is a non-zero probability for $x<0$ that does not exist in reality since $x$ is the concentration of the TF-A monomer. 
Still, the purpose of this numerical experiment is to show that the A-VSGA can approximate PDFs of systems having rational functions. 
Therefore, the existence of non-zero probability in the negative region is not relevant for our purpose.
For all the cases we have tested, the A-VSGA provides close PDFs as compared to those obtained analytically. 
In FIG.~\ref{fig:station_Hill}, the dashed lines describe each of the Gaussian basis functions that comprise the PDFs of the A-VSGA. Two bases each approximates one of the two peaks, while another basis is located at the center of the two peaks, reducing the error of the PDFs. 
Still, we see a small deviation from the analytical calculations in the A-VSGA results, which is the limitation of approximating PDFs with $N_B=3$ basis functions. 
For all parameter settings we have put in this experiment, the maximum feasible value of $N_B$ is 3. When the noise intensity approaches 0.1, it is difficult to find valid initial values. The nonorthogonality of multiple Gaussian distributions causes numerical instability, which also occurs in the original VSGA. 

\begin{figure}[t]
\centering
\includegraphics[width=5.5cm]{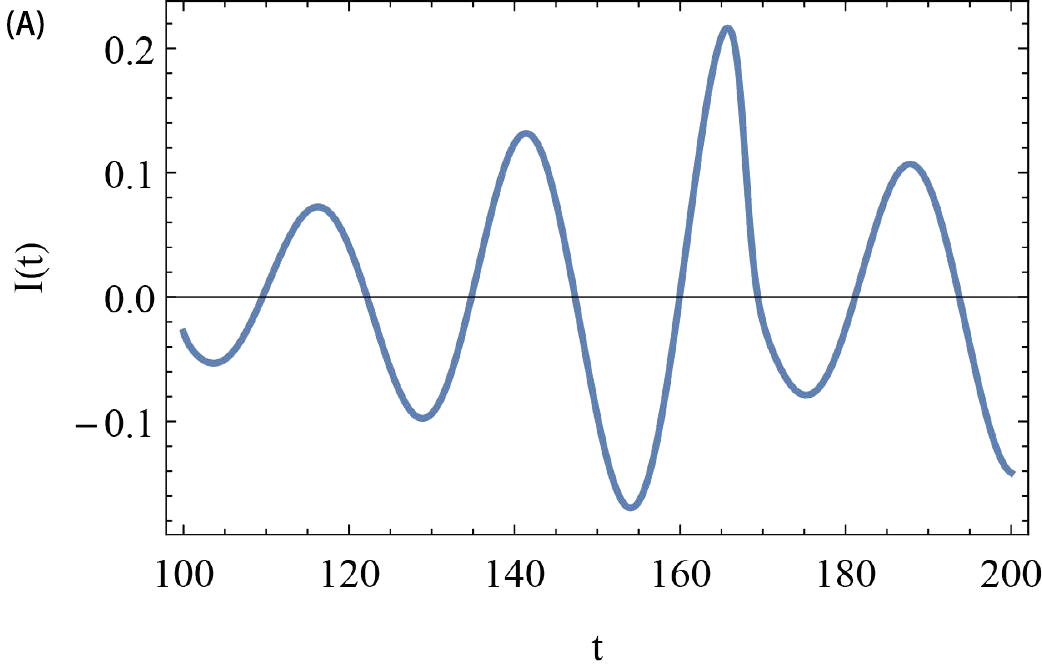}
\includegraphics[width=5.5cm]{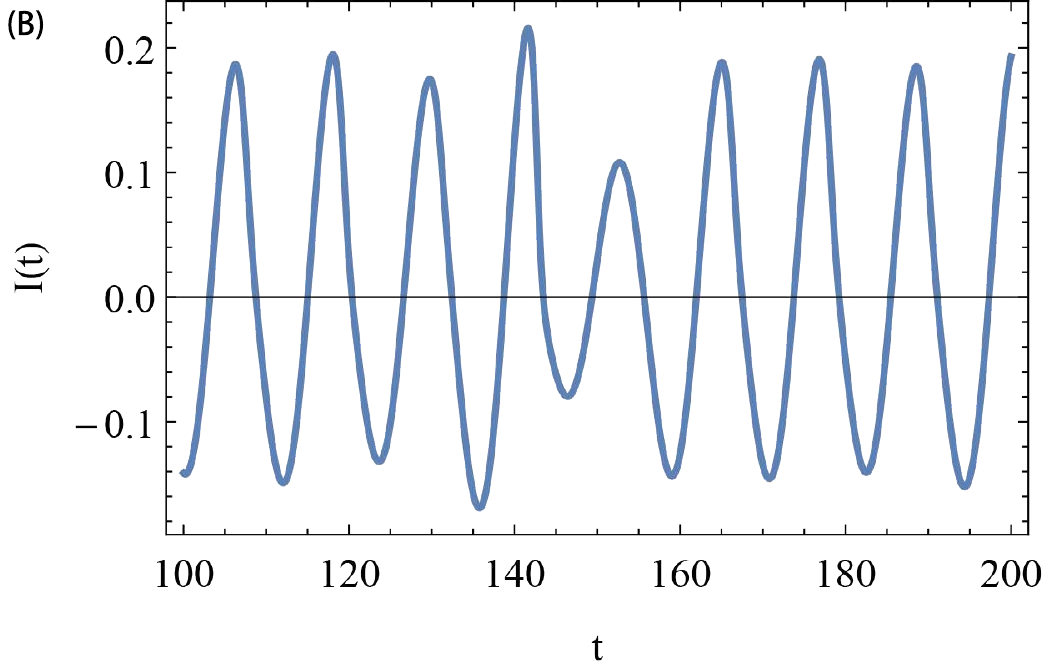}
\caption{Chaotic input signals. The parameters are (A) $\alpha=0.02$ and $\omega=0.25$ and (B) $\alpha=0.02$ and $\omega=0.5$.}
\label{fig:chaotic signals}
\end{figure}

Stochastic processes with external forces are non-stationary processes. For systems driven by periodic forces, time-dependent solutions are often represented as a Fourier series expansion \cite{jung1993periodic}. However, Fourier series expansion cannot be adopted to solving Langevin equations with chaotic external forces. In the dynamic case, we add a chaotic input signal to the system to show that the A-VSGA can be used for such systems. Consider a R\"{o}ssler oscillator:
\begin{align}
    \frac{ds_1}{dt}&=-s_2-s_3,\label{eq:s1}\\
    \frac{ds_2}{dt}&=s_1+c_1s_2,\label{eq:s2}\\
    \frac{ds_3}{dt}&=-c_2+s_3(s_1-c_3),\label{eq:s3}
\end{align}
where we use $c_1=0.2, c_2=0.2$, and $c_3=5.7$. The chaotic input signal is defined as follows
\begin{equation}
    I(t)=\alpha s_1(\omega t),\label{eq:input}
\end{equation}
where $\alpha$ is the input strength and $\omega$ is the reciprocal of the time scale. FIG.~\ref{fig:chaotic signals} shows trajectories of aperiodic input signals $I(t)$. The parameter settings are $\alpha=0.02$ and $\omega=0.25$ (FIG.~\ref{fig:chaotic signals}(A)) and $\alpha=0.02$ and $\omega=0.5$ (FIG.~\ref{fig:chaotic signals}(B)).

\begin{figure}[t]
\centering
\includegraphics[width=0.325\linewidth]{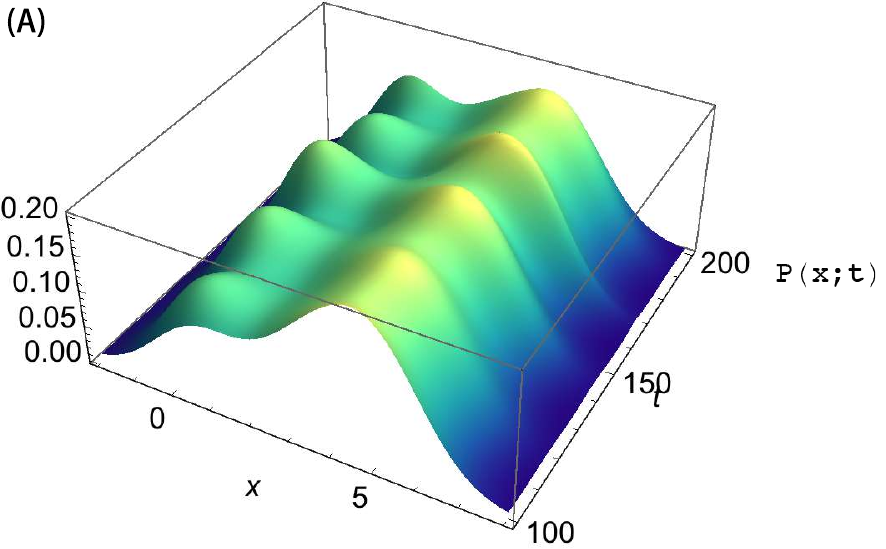}
\includegraphics[width=0.325\linewidth]{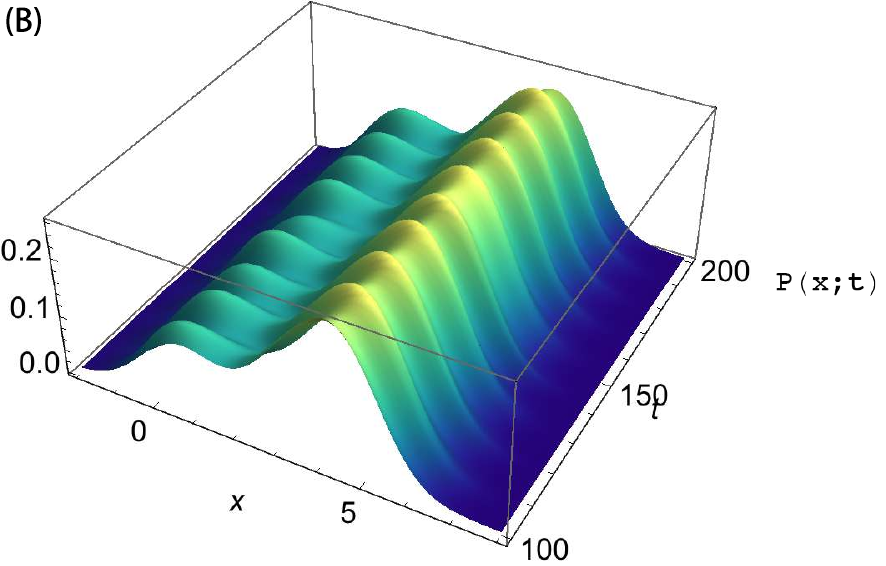}
\includegraphics[width=0.325\linewidth]{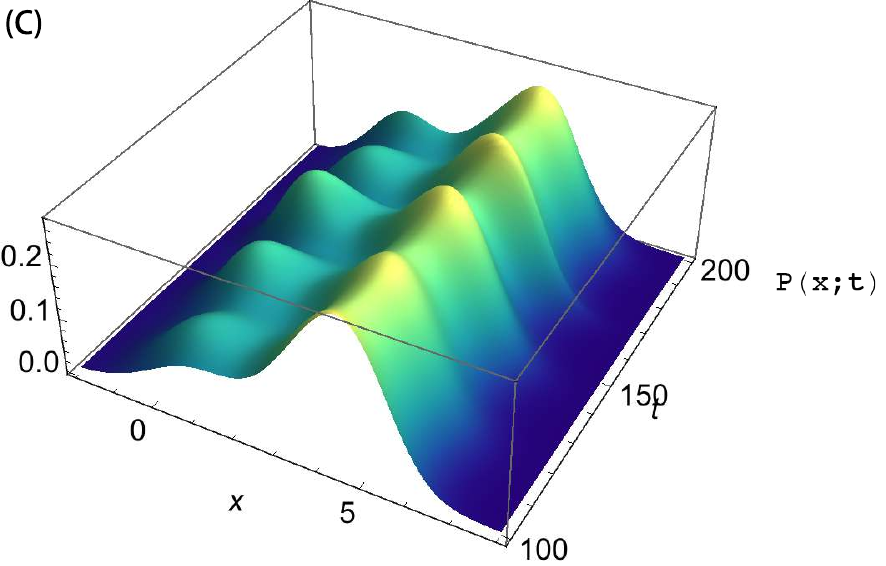}
\caption{(Color online) Dynamical PDFs, $P(x;t)$, of a gene regulatory system obtained by the A-VSGA with $N_B=3$ for (A) $D=1.0$, $\alpha=0.02$, and $\omega=0.25$; (B) $D=0.5$, $\alpha=0.02$, and $\omega=0.5$; and (C) $D=0.5$, $\alpha=0.02$, and $\omega=0.25$.}
\label{fig:Hill_3D}
\end{figure}

\begin{figure}[t]
\centering
\includegraphics[width=0.325\linewidth]{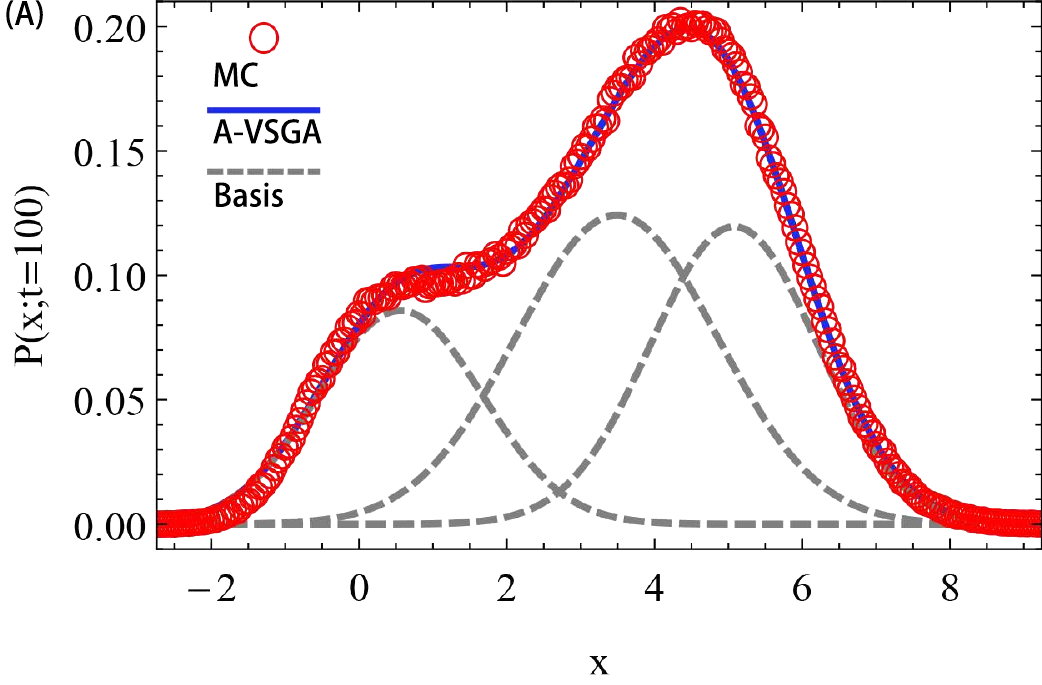}
\includegraphics[width=0.325\linewidth]{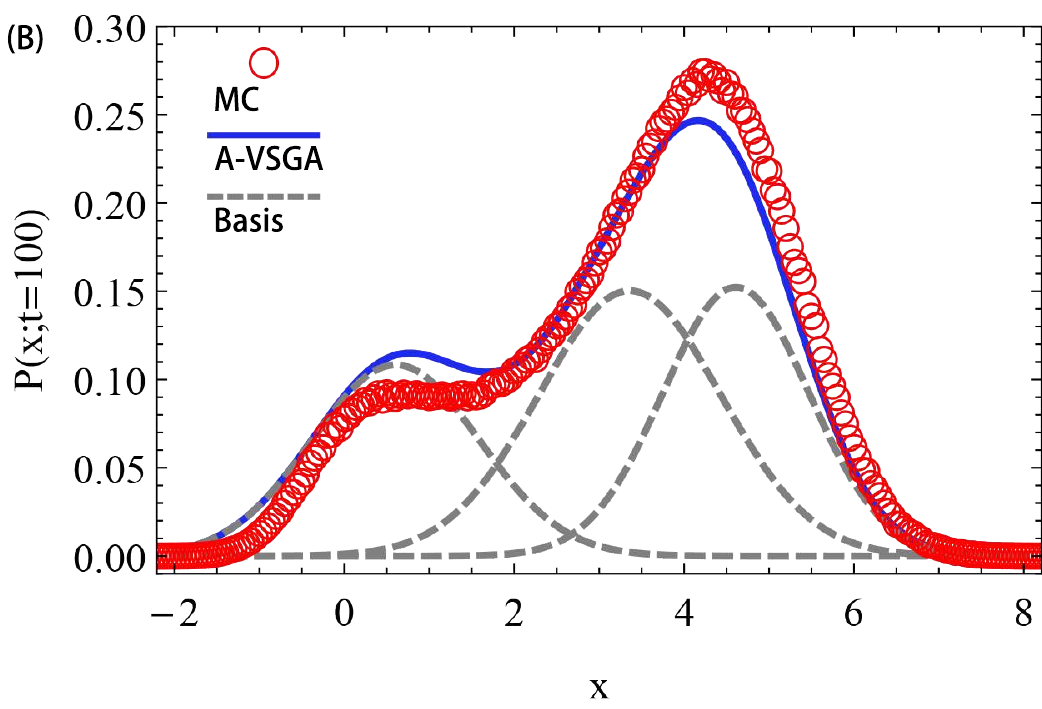}
\includegraphics[width=0.325\linewidth]{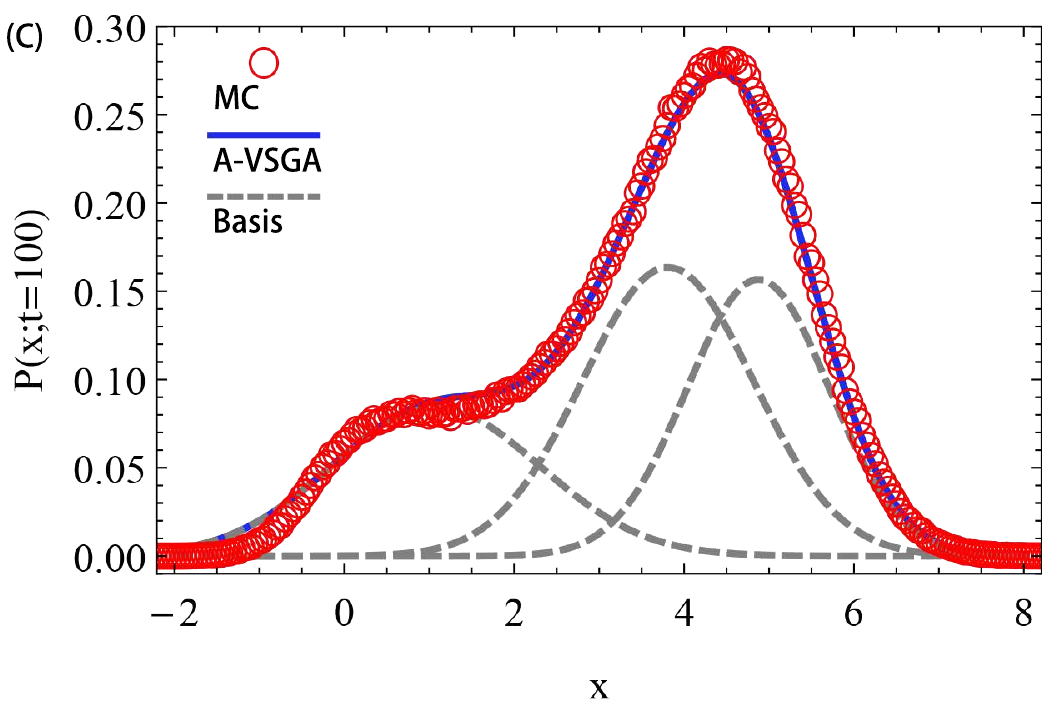}
\caption{(Color online) PDFs, $P(x;t)$, of a gene regulatory system at time $t=100$ obtained by MC simulations (circles) and the A-VSGA (solid line). The Gaussian basis functions are shown by the dashed lines. The parameter settings are (A) $D=1.0$, $\alpha=0.02$, and $\omega=0.25$, (B) $D=0.5$, $\alpha=0.02$, and $\omega=0.5$, and (C) $D=0.5$, $\alpha=0.02$, and $\omega=0.25$.}
\label{fig:Hill_MC}
\end{figure}

FIG.~\ref{fig:Hill_3D} shows the PDFs as functions of $t$ and $x$, which are calculated by the A-VSGA with the parameter settings $D=1.0$, $\alpha=0.02$, and $\omega=0.25$ (FIG.~\ref{fig:Hill_3D}(A)); $D=0.5$, $\alpha=0.02$, and $\omega=0.5$ (FIG.~\ref{fig:Hill_3D}(B)); and $D=0.5$, $\alpha=0.02$, and $\omega=0.25$ (FIG.~\ref{fig:Hill_3D}(C)).  
For non-linear systems driven by time-dependent external forces, there are no analytic solutions for Langevin equations. To investigate the accuracy of the $P(x;t)$ obtained by the A-VSGA, we performed MC simulations and compared the results of our method with that of MC simulations at a specific time. FIG.~\ref{fig:Hill_MC} shows the PDFs calculated by the A-VSGA (solid line), MC simulations (circles), and basis functions (dashed line) at time $t=100$. The parameter settings of FIG.~\ref{fig:Hill_MC} are the same as those for the data shown in FIG.~\ref{fig:Hill_3D}. 
When $\omega=0.25$ (FIG.~\ref{fig:Hill_MC}(A) and (C)), the PDFs of the A-VSGA agree very well with the results of MC simulations. The time-dependent solutions can be well approximated by 3 basis functions.
In FIG.~\ref{fig:Hill_MC}(B), the performance of the A-VSGA is not ideal. There is deviation between the curves around the peaks. The base near the left peak ($y=0$) is higher than the PDF obtained by the MC simulations, and the superposition of the two bases, near the right peak ($y=4.5$), is slightly lower than the solution obtained by the MC simulations. The results of our method would likely be more accurate in this case if more basis functions could be used.

\subsection{Soft Bistable System}

By using the Pad\'e approximant, several non-linear functions that cannot be approximated by the Taylor series expansion can be approximated by rational polynomial functions. Then, we can use the A-VSGA to obtain their time-dependent solutions. Here, we consider a soft bistable system \cite{longtin1994bistability}. The Langevin equation is
\begin{equation}
    \frac{dx}{dt}=-\frac{x}{2}+\tanh{x}+I(t)+\sqrt{D}\xi(t),\label{eq:soft bistable}
\end{equation}
where $I(t)$ is an input signal, $D$ is the noise intensity, and $\xi(t)$ is white Gaussian noise. The drift term is a hyperbolic function $\tanh{x}$ and not a polynomial.

\begin{figure}[t]
\centering
\includegraphics[width=6.5cm]{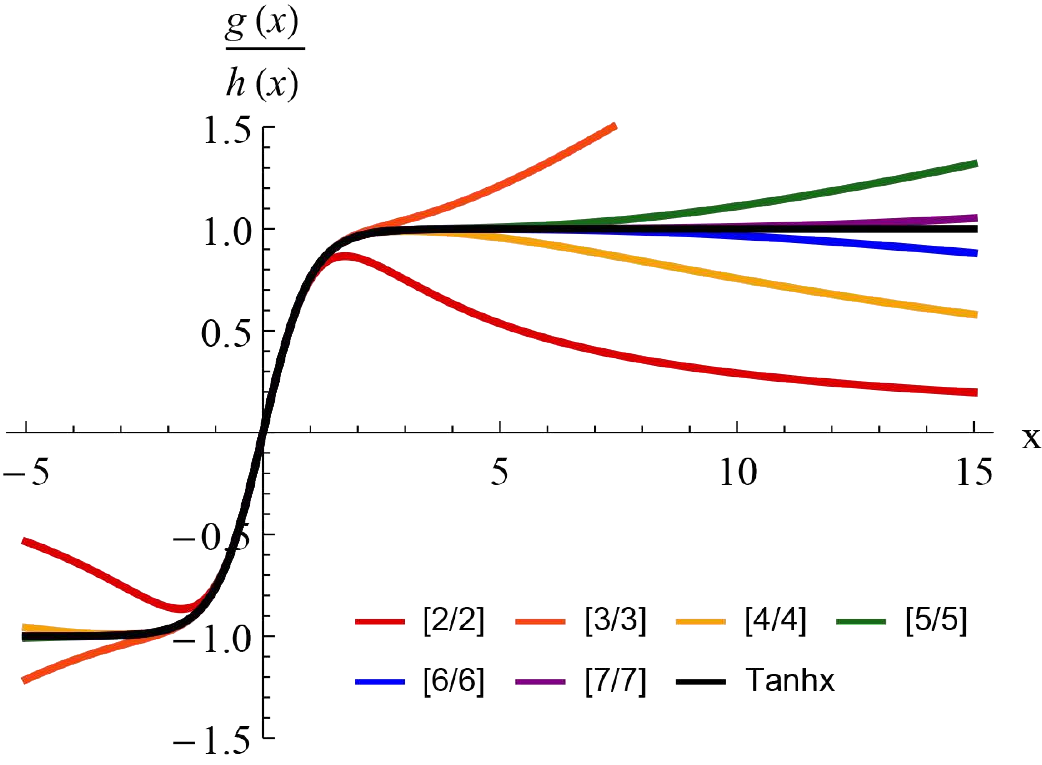}
\caption{$\tanh(x)$ (black line) and Pad\'e approximants of $\tanh(x)$ (colored lines) with different order $[m/n]$. }
\label{fig:pade of order[m/n]}
\end{figure}

The Pad\'e approximant can provide a drift function that is very close to $\tanh(x)$, whereas the Taylor series expansion of the original function diverges due to a pole at $|x|=\pi/2$. When we use the Pad\'e approximant to approximate the original function, we should pay attention to the following points: (1) the denominator of the approximant function should not have any poles, since poles lead the the A-VSGA to fail, (2) the Pad\'e approximant of the drift term should approach $\infty$ for $x\to \pm \infty$, which may not be satisfied for some order $[m/n]$ (as FIG.\ref{fig:pade of order[m/n]} shows, only odd order approximant of $\tanh{x}$ satisfy condition(2)), and (3) a high order $[m/n]$ may cause unavailability of the initial conditions and result in an increase in the calculation time. FIG.\ref{fig:pade of order[m/n]} shows several Pad\'e approximant of $\tanh(x)$ with different order $[m/n]$. In this case, when the order is odd, the approximant function satisfies both conditions (1) and (2). When the approximant order is $[5/5]$ or higher, there are more than 20 of the highest order of  moment functions, which causes extremely long computation and numerical instability in the A-VSGA. Here, for the given function, we chose a Pad\'e approximant of order $[3/3]$, 
\begin{equation}
    \frac{g(x)}{h(x)}=\frac{x+\frac{x^{3}}{15}}{1+\frac{2x^{2}}{5}}.
\end{equation}
Thus, the corresponding FPE operator $\widehat{L}(x;t)$ is given by,
\begin{equation}
    \widehat{L}(x;t)=-\frac{\partial}{\partial x}\left[\frac{x+\frac{x^{3}}{15}}{1+\frac{2x^{2}}{5}}-\frac{x}{2}+I(t)\right]+D\frac{\partial^{2}}{\partial x^{2}}.\label{eq:FPE_op_soft bistable}
\end{equation}
Substituting Eq.~\eqref{eq:FPE_op_soft bistable} into Eq.~\eqref{eq:ratpoly_coupled_equations}, we again have $3N_B+1$ constraints, including a normalization condition.

\begin{figure}[t]
\centering
\includegraphics[width=0.325\linewidth]{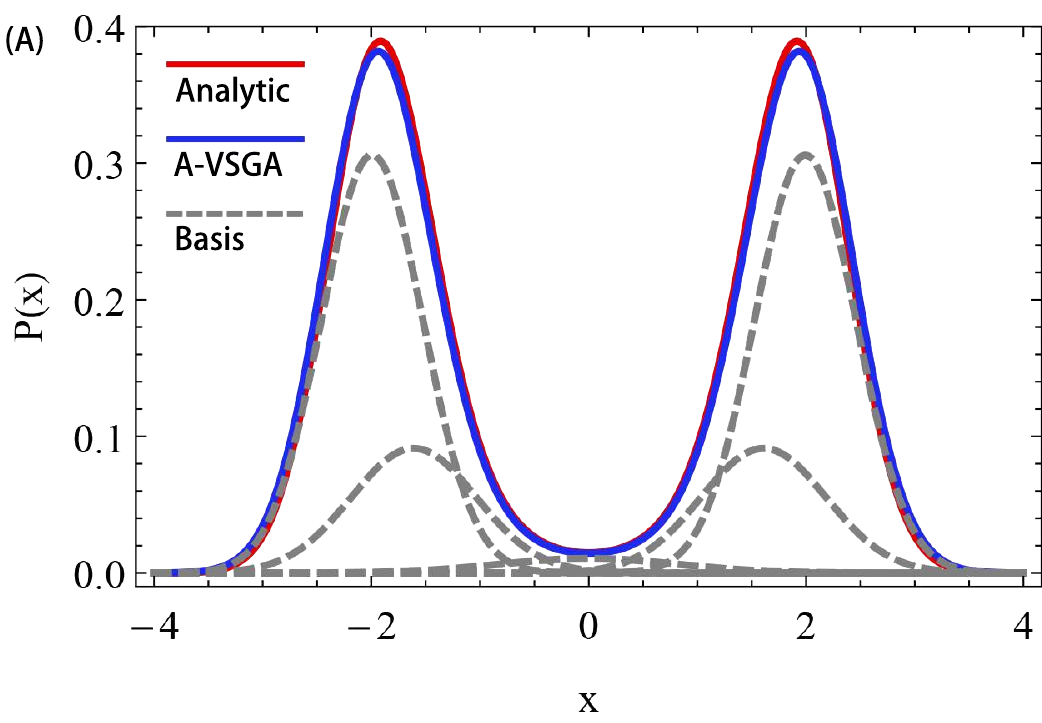}\label{main:soft_a}
\includegraphics[width=0.325\linewidth]{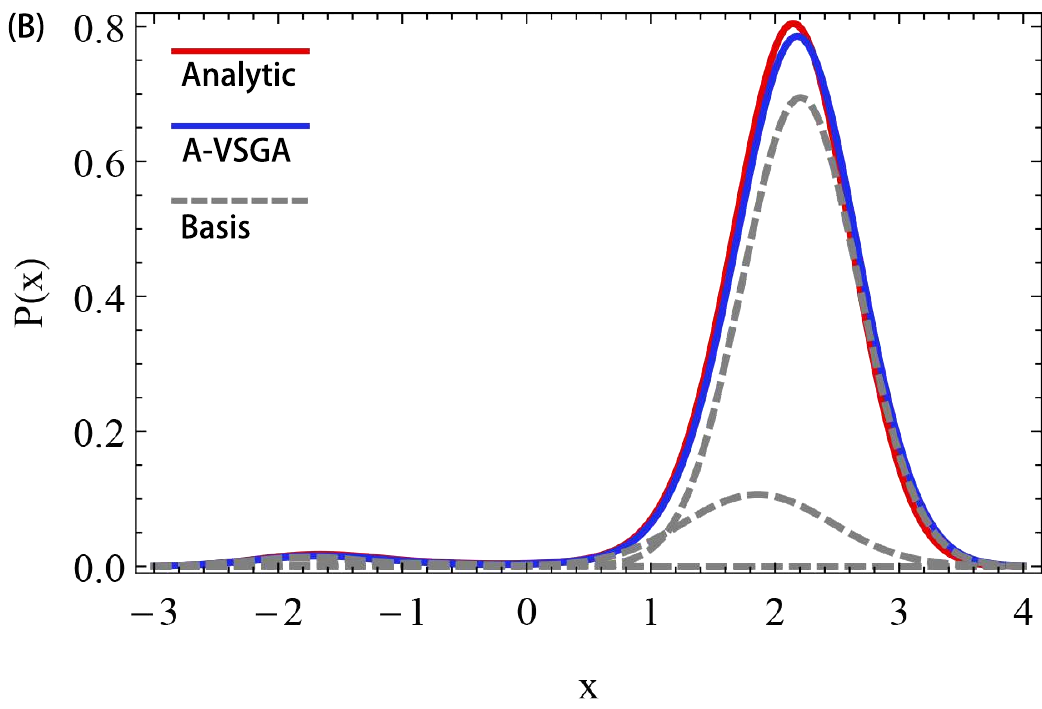}\label{main:soft_b}
\includegraphics[width=0.325\linewidth]{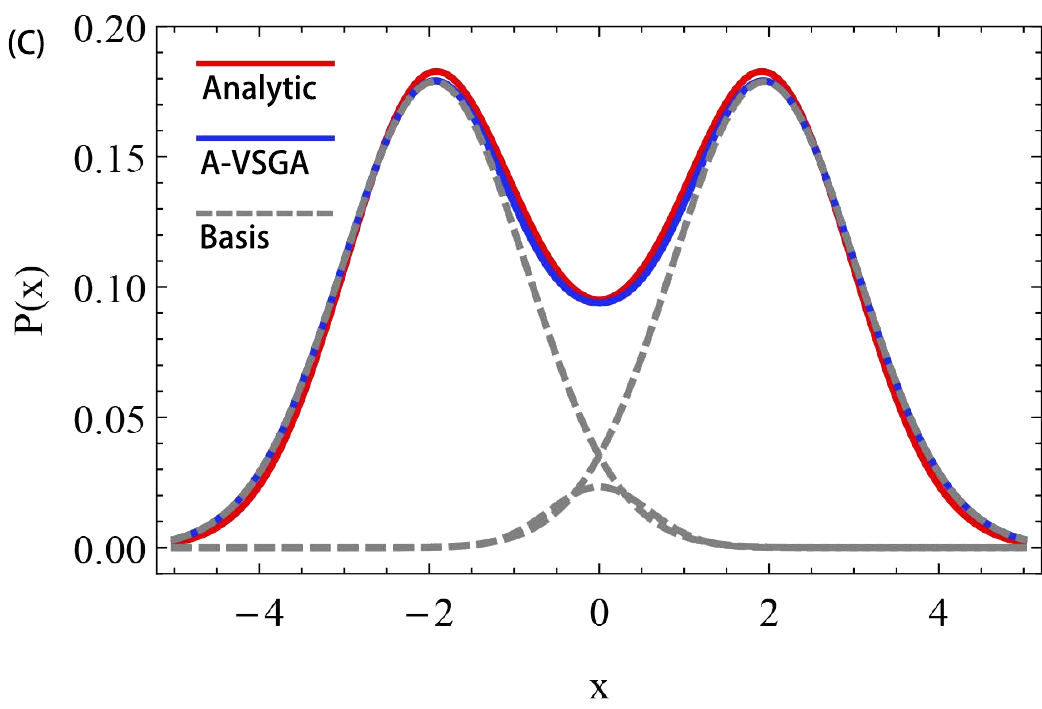}\label{main:soft_c}
\caption{(Color online) Stationary PDFs of the soft bistable potential obtained  by analytic calculation of Eq.~\eqref{eq:PDF_analytic} (red solid line) and the A-VSGA (blue solid line) for (A) $D=0.1$, $I(t)=0$, and $N_B=5$; (B) $D=0.1$, $I(t)=0.1$, and $N_B=5$; and (C) $D=0.5$, $I(t)=0$, and $N_B=3$. The grey dashed lines represent the basis functions.}
\label{fig:soft_station}
\end{figure}

Again, we first consider a stationary solution of the soft bistable model where the analytical PDFs can be obtained by the method described in \cite{er1998multi}.
FIG.~\ref{fig:soft_station} shows the results of the our approach (blue line) and the analytic solutions (red line) obtained by Eq.~\eqref{eq:PDF_analytic} for $D=0.1$, $I(t)=0$, and $N_B=5$ (FIG.~\ref{fig:soft_station}(A)); $D=0.1$, $I(t)=0.1$, and $N_B=5$ (FIG.~\ref{fig:soft_station}(B)); and $D=0.5$, $I(t)=0$, and $N_B=3$ (FIG.~\ref{fig:soft_station}(C)). For the analytical solutions, the potential function $U(x)=-\int (\tanh x - x/2)dx = {x}^{2}/4-\ln  \left( \cosh \left( x \right)  \right)$ is inserted in Eq.~\eqref{eq:PDF_analytic}. 
For $D\textgreater0.5$, the PDFs can be approximated by 5 basis functions (dashed line). Two bases each are located near the deterministic stable steady states ($y=\pm2$), and one is near the deterministic unstable steady state ($y=0$). The maximum feasible value of $N_B$ is clearly larger than that in the previous experiment. The stationary PDFs obtained by the A-VSGA are accurate. A large number of basis functions significantly improves the accuracy of the results. The A-VSGA also shows very good agreement with the analytical solutions in panel FIG.~\ref{fig:soft_station}(C), though only 3 basis functions are used.
In this experiment, the highest order of moment function in Eq.~\eqref{eq:ratpoly_coupled_equations} is the same as that in the first experiment, the gene regulatory system. However, here the results are more accurate. The only difference between the two experiments is that the complexity of Eq.~\eqref{eq:ratpoly_coupled_equations} of the soft bistable system is lower than that of the gene regulatory system. So, we suggest that a complex Eq.~\eqref{eq:ratpoly_coupled_equations} may increase the calculation difficulty, thus resulting in extensive temporal cost.

\begin{figure}[t]
\centering
\includegraphics[width=0.325\linewidth]{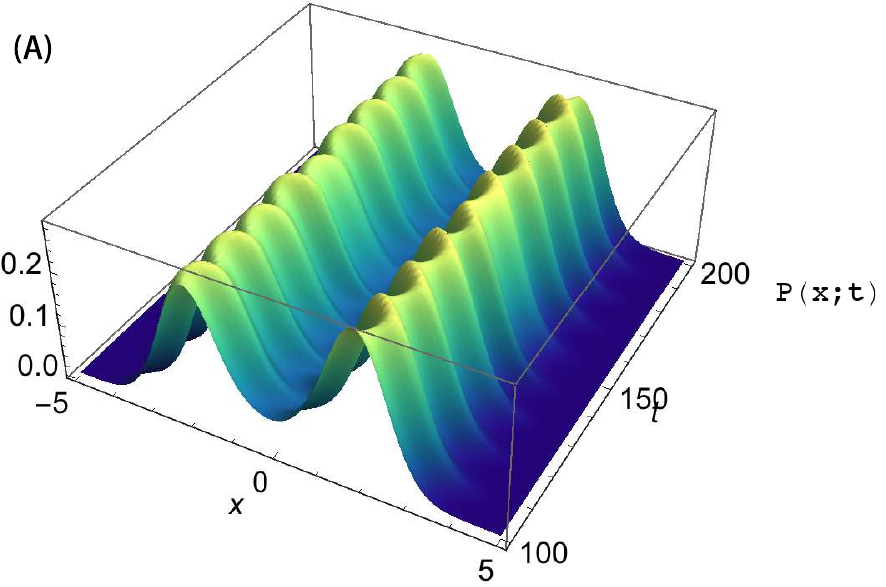}
\includegraphics[width=0.325\linewidth]{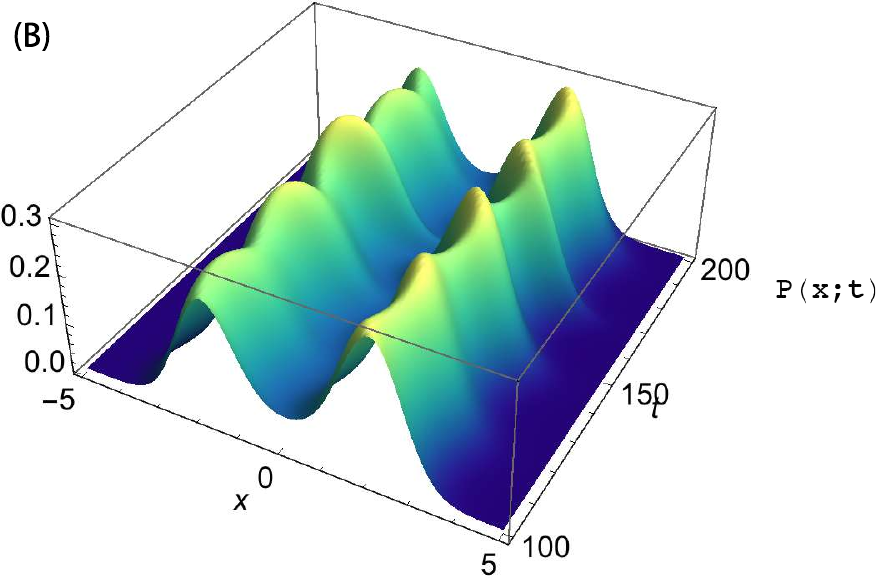}
\includegraphics[width=0.325\linewidth]{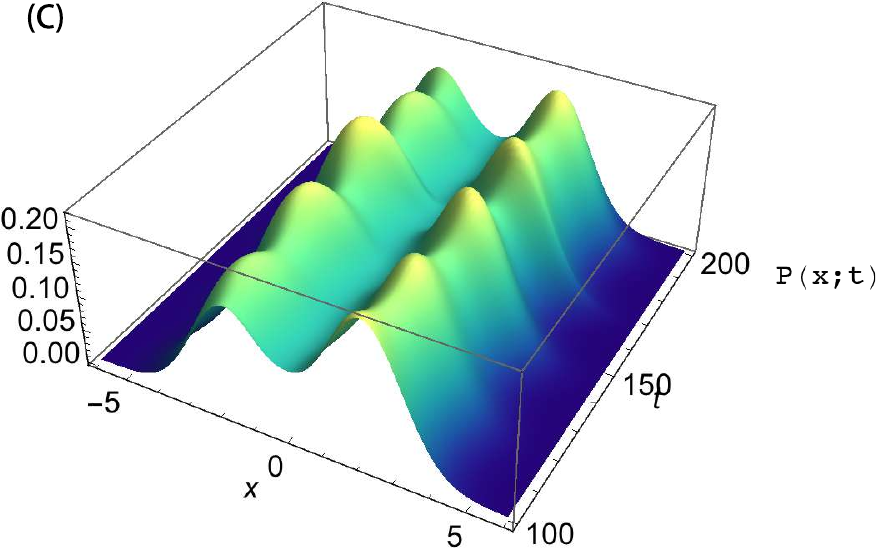}
\caption{(Color online) Dynamic PDFs $P(x;t)$ of the soft bistable potential obtained by the A-VSGA with (A) $D=0.2$, $\alpha=0.02$, $\omega=0.5$, and $N_B=4$; (B) $D=0.2$, $\alpha=0.02$, $\omega=0.25$, and $N_B=4$; and (C) $D=0.5$, $\alpha=0.02$, $\omega=0.25$, and $N_B=3$.}
\label{fig:tanh_3D}
\end{figure}

\begin{figure}[t]
\centering
\includegraphics[width=0.325\linewidth]{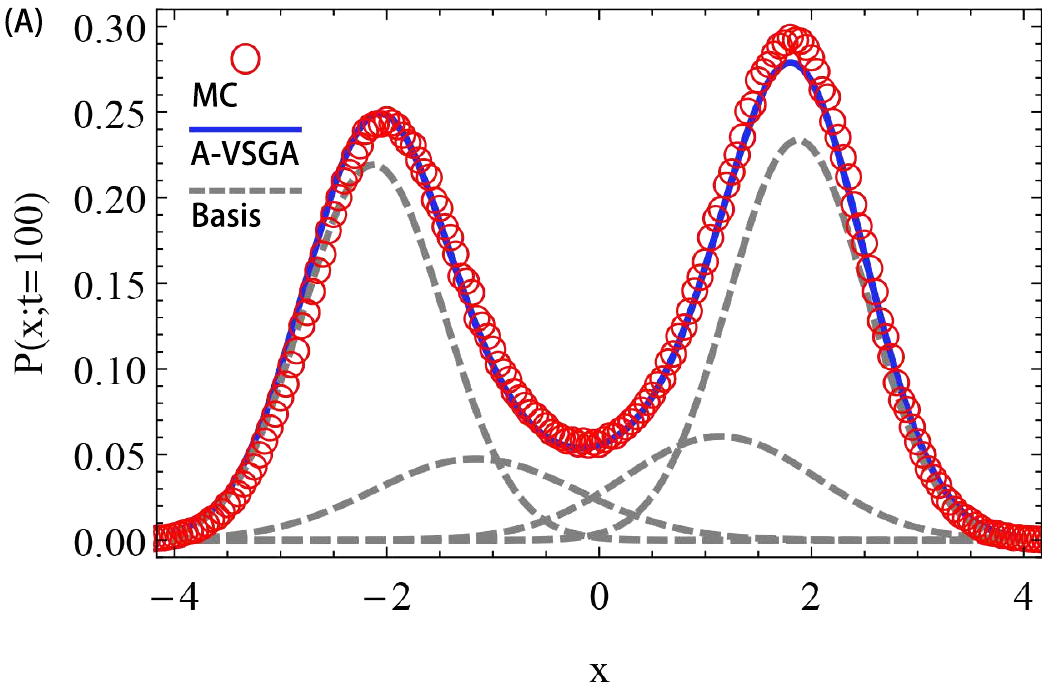}
\includegraphics[width=0.325\linewidth]{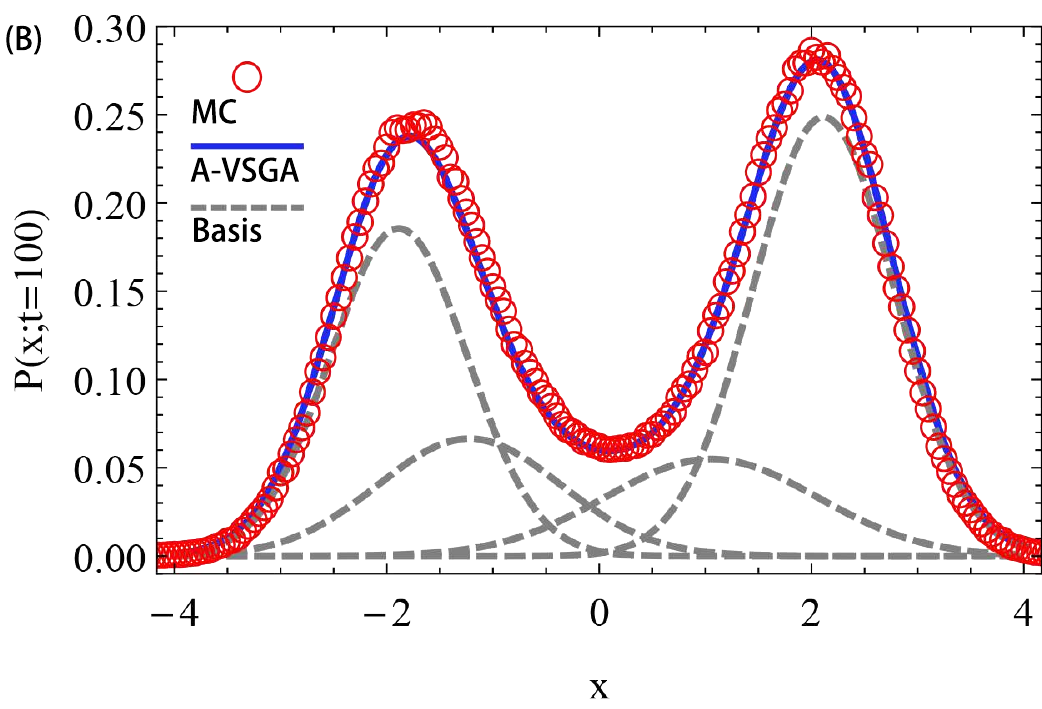}
\includegraphics[width=0.325\linewidth]{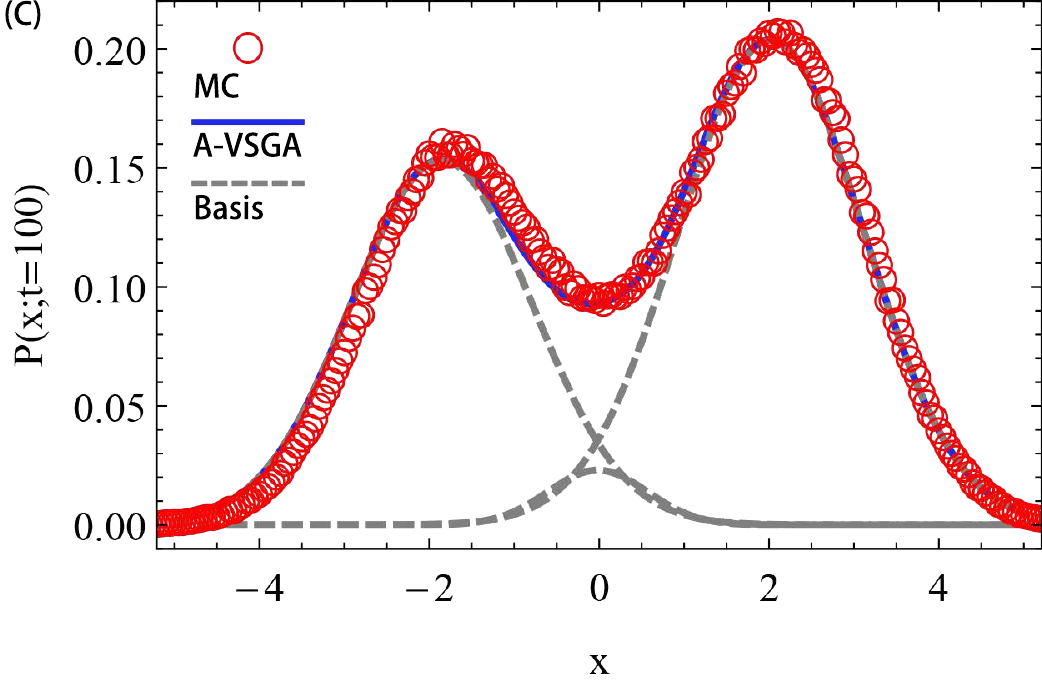}
\caption{(Color online) PDFs, $P(x;t)$, of the soft bistable potential at time $t=100$ obtained by MC simulations (circles) and A-VSGA (solid line). The Gaussian basis functions are shown by the dashed lines. The parameter settings are (A) $D=0.2$, $\alpha=0.02$, $\omega=0.5$, and $N_B=4$; (B) $D=0.2$, $\alpha=0.02$, $\omega=0.25$, and $N_B=4$; and (C) $D=0.5$, $\alpha=0.02$, $\omega=0.25$, and $N_B=3$.}
\label{fig:tanh_MC}
\end{figure}

Next we study the time-dependent solutions of a driven system. We add a chaotic input signal (defined by Eq.~\eqref{eq:input}) to the soft bistable system.
FIG.~\ref{fig:tanh_3D} shows the PDFs of the soft bistable system driven by the external input signal $I(t)$ with the following parameter settings: $D=0.2$, $\alpha=0.02$, and $\omega=0.5$ (FIG.~\ref{fig:tanh_3D}(A)); $D=0.2$, $\alpha=0.02$, and $\omega=0.25$ (FIG.~\ref{fig:tanh_3D}(B)); and $D=0.5$, $\alpha=0.02$, and $\omega=0.25$ (FIG.~\ref{fig:tanh_3D}(C)). 
FIG.~\ref{fig:tanh_MC} shows the time-dependent solutions, $P(x;t)$, at $t=100$, which are calculated by the A-VSGA (solid line) and MC simulations (circles) with the same parameter settings used for the PDFs of the soft bistable system (FIG.\ref{fig:tanh_3D}). 
When noise intensity $D=0.2$, the PDFs can be approximated by 4 Gaussian distributions (FIG.~\ref{fig:tanh_MC}(A) and (B)). Near each of the two peaks ($y=\pm2$) is a base, and at the center of each of the two peaks ($y=0$) there is a base.
For all cases we have tested, the results of the A-VSGA are in excellent agreement with results of the MC simulations. In FIG.~\ref{fig:tanh_MC} (C), the PDF is a superposition of 3 Gaussian distributions.
This verifies the reliability of the A-VSGA with respect to the PDFs at a specified time in dynamic case.

\section{Conclusion\label{sec:conclusion}}

We propose the A-VSGA for the time-dependent solutions of Langevin equations with rational or several other non-linear drift terms. Our approach is applied to two systems to show its reliability. Compared with the MC simulations, the A-VSGA is an efficient method to calculate time-dependent solutions without relying on stochastic approaches. The results of the soft bistable system in Sec.~\ref{sec:results} are satisfactory and agree well with the PDFs from the MC simulations.

Our approach uses multiple Gaussian distributions to approximate PDFs, which is the same as is done in the VSGA. A large number of basis functions that are not orthogonal prevents the calculation of the time evolution of the parameters. It can be seen from Sec.~\ref{sec:method} that soving Eq.~\eqref{eq:ratpoly_coupled_equations} in the A-VSGA is more complicated than solving Eq.~\eqref{eq:DAE_solution} in the VSGA. The complexity of Eq.~\eqref{eq:ratpoly_coupled_equations} increases the computation difficulty. 
Considering the two experiments carried out in Sec.~\ref{sec:results}, the maximum feasible value of $N_B$ of the soft bistable system is five, whereas no more than three basis functions are used in the gene regulatory system.
The reason is that Eq.~\eqref{eq:ratpoly_coupled_equations} in the gene regulatory system is more complex than that of the soft bistable system.
Moreover, an overly complex calculation in the gene regulatory system results in no valid initial values for the A-VSGA when the noise intensity is relatively small.
Thus, the complexity of Eq.~\eqref{eq:ratpoly_coupled_equations} is related to the time cost and difficulty of the calculation.

The A-VSGA can be used to obtain time-dependent solutions of Langevin equations whose drift terms are expressed by rational function or other forms such as a hyperbolic function. We have shown that the A-VSGA can effectively provide accurate approximation in non-equilibrium systems. This approach can be used for analyzing certain real-world problems such as enzymatic reactions, cell cycles, gene regulation, and a variety of ligand-receptor interaction problems.

\end{document}